# Graphene-Confined Growth of Few-Layer $PtSe_2$ via One-Step Selenization

Timea Ema Krajčovičová,[1] Peter Kotrusz,[1,2] Viera Skákalová,[1,2] Agáta Laurenčíková,[1] Igor Píš,[1,3] Martin Hulman,[1,2] Kimmo Mustonen[4] and Michaela Sojková,[1*]

[1]*Institute of Electrical Engineering, Slovak Academy of Sciences, Dúbravská cesta 9, 84104 Bratislava, Slovakia*

[2]*Danubia NanoTech, s.r.o., Dúbravská cesta 9, 84104 Bratislava, Slovakia*

[3]*Polymer Institute, Slovak Academy of Sciences, Dúbravská cesta 9, 845 41 Bratislava, Slovakia*

[4]*University of Vienna, Faculty of Physics, Boltzmanngasse 5, 1090 Vienna, Austria*

**Corresponding author:**

email: michaela.sojkova@savba.sk

**ABSTRACT:**

Platinum diselenide ($PtSe_2$), a member of the transition metal dichalcogenide (TMD) family, is a promising material for next-generation electronic and optoelectronic applications due to its thickness-dependent properties, transitioning from semimetallic in bulk to semiconducting behavior at the few-layer limit. However, the scalable fabrication of large-area ultrathin $PtSe_2$ films remains challenging, as conventional selenization of pre-deposited Pt layers often leads to undesired three-dimensional clustered growth. This work explores a novel strategy for promoting two-dimensional growth of $PtSe_2$ by employing a graphene-mediated confinement approach. The method involves coating a Pt layer with graphene oxide (GO), followed by annealing in a selenium atmosphere, which results in the simultaneous formation of $PtSe_2$ and the reduction of GO to graphene. The graphene overlayer acts as a physical confinement layer, suppressing vertical growth and enabling the synthesis of graphene-confined $PtSe_2$ heterostructures on centimetre-scale substrates and promoting lateral connectivity of $PtSe_2$ regions observed by STEM. Scanning transmission electron microscopy (STEM), including temperature-resolved in-situ atomic resolution measurements, was used to investigate the influence of selenization temperature on the growth mechanism and structural evolution of $PtSe_2$ films. The results show that graphene confinement stabilizes lateral growth and promotes the formation of few-layer films with improved structural ordering. Beyond $PtSe_2$, this graphene-assisted approach provides a general strategy for fabricating and stabilizing ultrathin TMD layers, including those prone to environmental degradation, providing a new outlook for enabling detailed studies of their intrinsic properties.

## INTRODUCTION

The fundamental quality of a 2D material is its scalability to form atomically thin layers that provide near-ideal electrostatic gate control, even at nanometer-scale channel lengths. Among several 2D materials, transition metal dichalcogenides (TMDs) are of particular interest, standing out for their facile and diverse synthesis methods and transferability to a variety of substrates, including flexible and transparent ones.[1] Materials of the TMD family can be engineered with diverse properties through various combinations of metals and chalcogenides, spanning not only semiconductors ($WS_2$, $MoS_2$) but also metals and semimetals ($VSe_2$, $WTe_2$, $TiSe_2$), superconductors ($NbSe_2$, $TaS_2$), and insulators ($HfS_2$).[2]

A key characteristic of all 2D materials is their unique electronic behavior, which differs greatly from that of their bulk counterparts. Structurally, bulk TMDs resemble graphite, consisting of planes weakly bound by van der Waals forces, easily exfoliated into monolayers. In monolayer form, some TMDs ($MoS_2$, $WS_2$) shift from indirect to direct band gap semiconductors, with the valence and conduction band edges located at the $K$ points of the Brillouin zone, as opposed to bulk. This transition greatly enhances their optical and electronic properties, making them highly sought after in current research.[3,4]

Among TMDs, platinum diselenide is particularly notable, because it combines several properties that are rarely found together in other 2D materials. $PtSe_2$ is a group-10 transition-metal dichalcogenide that crystallizes in the octahedrally coordinated 1T structure. In contrast to widely studied 2H-phase TMDs, such as $MoS_2$ and $WS_2$, which are typically semiconducting, bulk $PtSe_2$ exhibits semimetallic electronic behavior. Its electronic structure is strongly thickness dependent: reducing the thickness to the ultrathin regime opens a band gap, and monolayer as well as selected few-layer $PtSe_2$ have been reported to exhibit semiconducting behavior, whereas thicker $PtSe_2$ progressively approaches the semimetallic bulk state. [5,6] This thickness dependence makes $PtSe_2$ of interest for electronic and optoelectronic applications.[7]

The transport properties of $PtSe_2$ reported in the literature vary besides layer thickness also with crystallinity, defect density, substrate interaction, contact configuration, and the measurement method. Although some theoretical studies predict room-temperature phonon-limited carrier mobilities above 1000 $cm^2/V.s$ for $PtSe_2$,[8,9] experimentally measured mobilities are generally considerably lower and strongly depend on material quality and measurement configuration. Reported room-temperature values range from 13–24 $cm^2/V.s$ for $PtSe_2$ films prepared by thermally assisted conversion [10,11] to 210 $cm^2/V.s$ for mechanically exfoliated flakes[12] and more than 300 $cm^2/V.s$ for CVD-grown nanosheets.[13]

Generally, two main strategies are employed to fabricate 2D materials. 2D TMD layers can be separated from bulk through exfoliation methods such as mechanical or chemical, in a so-called top-down approach. In a bottom-up approach, materials are grown directly on target substrates from elemental precursors. This includes methods such as chemical vapor deposition, molecular beam epitaxy, and atomic layer deposition. Top-down exfoliation methods, including mechanical and liquid-phase exfoliation, provide access to high-quality $PtSe_2$ flakes,[14] although their lateral dimensions, thickness uniformity, and scalability are limited. In contrast, bottom-up approaches enable direct growth on target substrates and offer greater potential for controlled large-area deposition and integration into device-relevant architectures.[15] Table 1 presents a short summary of the advantages and disadvantages of common growth methods of TMD materials.

When aiming for an intersection between large area coverage of TMD films and simple instrumentation, emerging as a very promising method for thin film growth is a variation of vapor deposition techniques - thermally assisted conversion (TAC). The principle of this method is the annealing of pre-deposited metal films in chalcogen vapor, typically by sputtering or evaporation. This method is a relatively fast and simple process and does not require complicated setups with multiple temperature zones or vacuum conditions. The thickness of the resulting TMD thin film is tunable by adjusting the thickness of the metal deposited on the substrate. Therefore, this method is of interest because of its potential for scalable $PtSe_2$ synthesis. Nevertheless, monolayer growth using this method remains largely unexplored.[11,16]

Other than the influence of different synthesis methods, another important variable in producing 2D material layers is the choice of substrate,[17] or rather, a material capable of acting as a stabilizing template.[18] Moon et al.[19] have coined the term "hypotaxy" and achieved remarkable results in growing single-crystal TMD materials ($MoS_2$, $MoSe_2$, $WS_2$, and $WSe_2$) of wafer-scale dimensions. In short, a thin metal film is first deposited on a substrate, then covered with a layer of graphene. During high-temperature chalcogenization at 1000 °C, nanopores form in the graphene, allowing chalcogen atoms to diffuse through and react with the metal. This leads to the formation of TMD nuclei which grow and merge into a continuous, layered TMD film. The key principle of the hypotaxy process described in this work is that the resulting TMD layers inherit their crystallographic alignment from the graphene template rather than from the underlying substrate. It is demonstrated that the grain structure of the grown material directly reflects that of the graphene template, meaning when single-crystal graphene was used, the resulting TMD film also formed a large area single crystal.[19]

Similarly, Mustonen et al.[20] have presented a novel approach for stabilizing monolayers of β-phase cuprous iodide (CuI), a material typically only occurring in layered form in a narrow range of elevated temperatures. This work presents a method of graphene encapsulation, making it possible to observe and verify the properties of a single atomic layer β-CuI under ambient conditions when sandwiched between graphene sheets. The authors demonstrate that the stability of the cuprous iodide monolayers was dependent entirely on the confinement and pressure provided by the graphene encapsulant and without it, were quick to degrade.[20]

It is clear there is much to explore regarding the influence of graphene on the study of 2D materials. Building on these findings, our work investigates the role of graphene both as a substrate and encapsulant of $PtSe_2$ in the form of heterostructures. Graphene displays superior mechanical strength, thermal and electrical conductivity and a high active surface area. A major limiting factor for its application in electronic devices is, however, a lack of bandgap,[21] and therefore, it may be better suited to take on a supportive role for materials with semiconducting properties. The existing literature on graphene-TMD heterostructures discusses their applicability as e. g. photodetectors, field effect transistors, and sensors.[22] The construction of these heterostructures typically relies on the manual stacking of exfoliated graphene sheets; however, such an approach lacks scalability and opens the process to contamination or mechanical damage.[23]

Our work proposes a simple formation of a $PtSe_2$-graphene heterostructures. The aim is to increase the crystal quality of the ultra-thin TMD layers and to protect them from environmental degradation. The role of graphene is crucial in that by reducing the space for 3D growth, it promotes horizontal growth of TMD grains leading to a more uniform and lateral coverage on the substrate.

## EXPERIMENTAL

$PtSe_2$ films were prepared by selenization of pre-deposited platinum layers, the TAC method. Typically, this method utilizes two temperature zones – the chalcogen powder source is evaporated around temperatures just above its melting point, but the metal substrate is heated to higher temperatures. Our experiments use a simplified approach – both the substrate and the chalcogen powder were placed at the same position in the center of a one-zone furnace. This makes it possible to use an affordable vacuum-free experimental setup.

Heterostructures of platinum diselenide and graphene were directly fabricated on custom made silicon dioxide/nitride TEM grids (Silson Ltd., UK) as substrates (Fig. 1). The grid has a diameter of 3 mm and contains nine individually supported, 1 μm thick, freestanding silicon nitride windows, each of 90 × 90 μm$^2$ area. Each window is perforated with an array of circular holes, each 3 μm in diameter, over which the samples are suspended, as shown in Figure 1. The grid was first covered by graphene oxide from aqueous solution by a dip coating and drying process, to support deposition of platinum. The graphene oxide (GO) was thermally reduced to reduced graphene oxide (rGO) by annealing at 300 °C. Next, a thin layer of platinum (0.5 nm) was deposited on the rGO surface by two different deposition techniques – magnetron sputtering and thermal evaporation. RF magnetron sputtering was performed in the Ar atmosphere ($10^{-3}$ mbar) from a platinum target at room temperature with the RF power of 25 W. AJA Orion 8E hybrid system was used for the e-beam evaporation of Pt layers. One type of sample was left as is, while the other type was once more covered by GO from aqueous solution. This process created two types of arrangements: rGO/Pt and rGO/Pt/GO. Both samples were subjected to selenization, during which sputtered platinum films were transformed into $PtSe_2$ films. Parameters were the following: annealing temperature – 400 and 550 °C, annealing time – 30 min, heating rate – 25 °C min$^{-1}$, nitrogen flow rate – 100 sccm. Graphene oxide is permeable to selenium atoms due to its oxidized structure that introduces interlayer spacing, defects, and hydrophilic nanochannels, allowing them to pass through and interact with platinum atoms, however, the high temperatures of the selenization process simultaneously reduce graphene oxide into graphene, which becomes significantly less permeable to selenium vapor. The resulting samples have the horizontal configuration of rGO/$PtSe_2$ and rGO/$PtSe_2$/rGO.

In addition to the SiN TEM-grid samples used for STEM characterization, equivalent heterostructures were prepared on 1×1 cm$^2$ sapphire substrates for Raman and X-ray photoelectron spectroscopy analyses. Sapphire-supported rGO/Pt/GO and Pt/GO heterostructures were prepared to assess the influence of the bottom rGO layer on $PtSe_2$ formation. Since AFM measurements showed no appreciable difference between $PtSe_2$ films grown on bare sapphire and rGO-coated sapphire (Fig. S1-2), the bottom rGO layer was omitted from the sapphire-supported Pt/GO reference samples. Both heterostructures were subsequently subjected to the same selenization process used for the TEM-grid samples. Only magnetron sputtered platinum samples were used for this purpose.

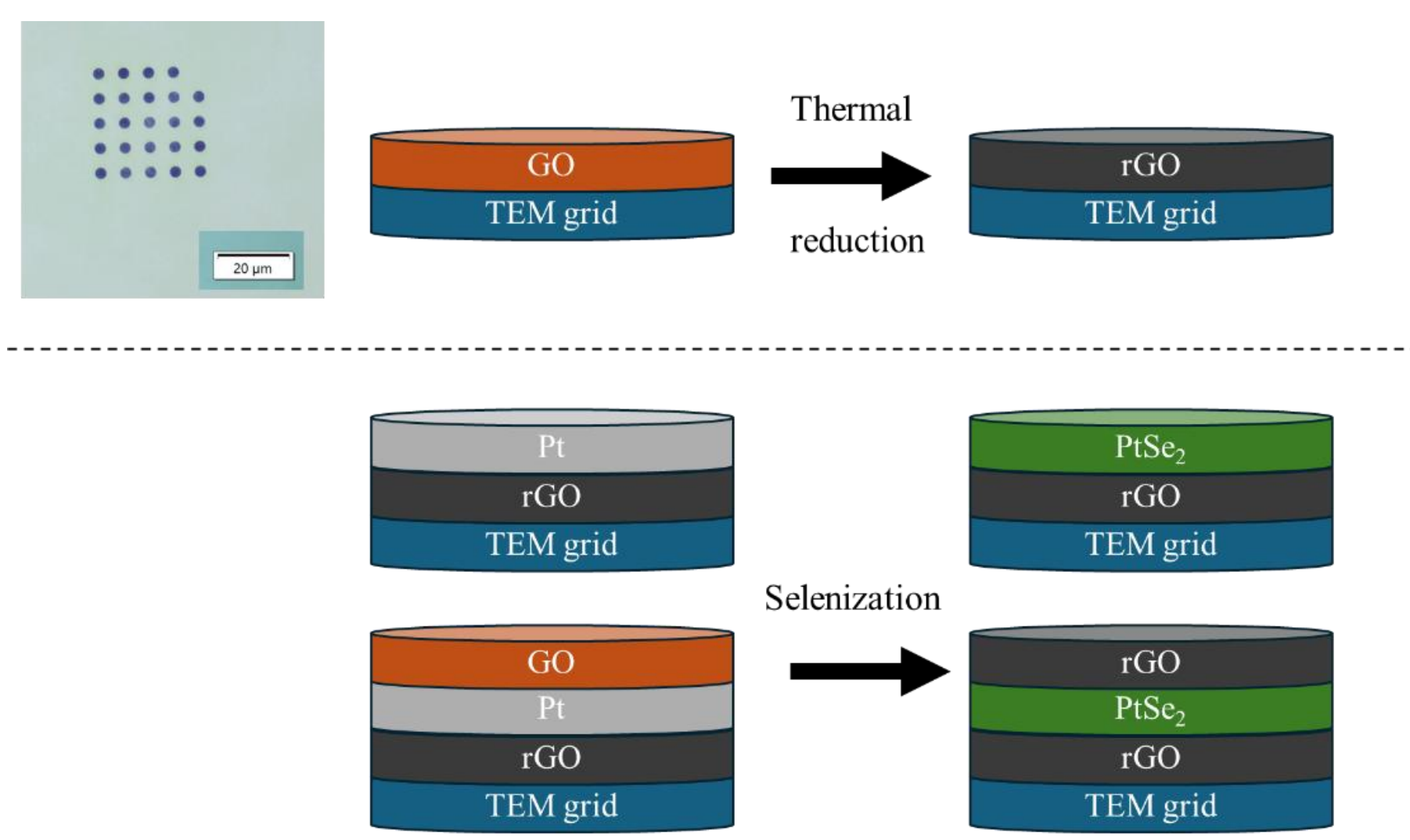

Figure 1 – Schematic representation of graphene-$PtSe_2$ heterostructures fabrication

For thermal stability experiments in Figure 9, the rGO/$PtSe_2$/rGO heterostructures were grown directly on Protochips Fusion heating holders (Protochips Ltd., USA), used for the in-situ heating experiments in the same manner as on the TEM grids.

The STEM data were acquired using an aberration-corrected Nion UltraSTEM100 scanning transmission electron microscope, operated at 60 keV electron energy, with the sample in ultrahigh vacuum conditions at a pressure of UHV ($\sim 10^{-9}$ mbar), using a high-angle annular dark field (HAADF) detector with a collection semi-angle of 80–300 mrad.

Raman spectra were acquired under ambient conditions using a confocal Raman microscope (alpha300 R, WITec, Germany) equipped with a 532 nm excitation laser. The laser power at the sample was limited to 1 mW to minimize laser-induced damage. Raman scattering was collected through a 50× objective lens (NA = 0.8) and detected using a Peltier-cooled EMCCD camera. Spectral dispersion was provided by an 1800 grooves/mm blazed grating, resulting in an overall spectral resolution of approximately 0.75 $cm^{-1}$.

XPS measurements were performed using a Thermo Scientific Nexsa G2 Surface Analysis System (Thermo Fisher Scientific, UK) equipped with a monochromatized Al Kα X-ray source (hν = 1486.68 eV) and a hemispherical analyser. An X-ray beam of 400 $\mu m^2$ was used. The spectra were acquired in constant-energy mode at a pass energy of 50 eV, with a total energy resolution of 0.6 eV at a take-off angle of 90° from the surface plane. Energy scale calibration was performed using the automated calibration routine and the internal Au, Ag, and Cu standards supplied with the NEXSA G2 system. Se 3d and C 1s core-level spectra were peak-fitted in KherveFitting [24] using Voigt profiles and a Shirley background. Pt 4f and C 1s $sp^2$ components were modeled with asymmetric Gaussian convoluted Doniach–Sunjic functions, with widths and asymmetry parameters determined from reference spectra (uncoated $PtSe_2$ sample, sputter-cleaned Pt metal, and reduced graphene oxide) and applied as constraints. Atomic concentrations (at. %) were calculated from peak areas using the relative sensitivity factors provided in the Avantage software package (Thermo Scientific).

## RESULTS AND DISCUSSION

The selenization temperature was selected based on both the known temperature dependence of $PtSe_2$ formation from pre-deposited Pt and the thermal tolerance of the GO-derived confinement architecture. In our earlier one-zone selenization studies, increasing the conversion temperature enhanced $PtSe_2$ crystallinity and promoted the formation of larger crystalline domains, whereas lower temperatures produced smaller and less ordered features.[11,25] However, the temperature cannot be increased without limit in the present heterostructure: when rGO/Pt/GO precursor stacks were processed at 600 °C, the heterostructure peeled off the SiN TEM grid.

We therefore selected 550 °C for 30 min as the most reproducible processing condition that preserved the sandwich architecture while improving lateral grain growth and structural ordering compared with samples converted at 400 °C. This temperature should be regarded as a practical compromise for the specific rGO/Pt/GO configuration and one-zone furnace conditions used here, rather than as a universal optimum for $PtSe_2$ synthesis.

In the TAC method, the thickness of the final $PtSe_2$ material is dependent on the thickness of the starting platinum film. $PtSe_2$ displays thickness-dependent electronic properties; however, the thickness associated with semiconducting behavior can depend on film quality, stoichiometry, defect concentration, stacking configuration, strain, and substrate interactions.[10,26,27] Selenization of Pt also leads to a substantial thickness increase, which must be considered when targeting ultrathin $PtSe_2$ films. In our previous work, X-ray reflectivity measurements indicated that selenization of a nominally 0.5 nm Pt precursor layer yielded an area-averaged effective $PtSe_2$ thickness of approximately 3 nm, corresponding to about six $PtSe_2$ layers.[11] The reported thickness is included to demonstrate the extent of film expansion during Pt selenization, rather than to infer the electronic character of the resulting $PtSe_2$ layer. This behavior can be attributed to the pronounced volume expansion that occurs during the conversion of Pt to $PtSe_2$, as selenium incorporation significantly increases the unit-cell volume and overall film thickness. Thus, in the first step, the character of initial Pt layers was studied. At the nominal Pt thickness of 0.5 nm, both deposition methods produced discontinuous, island-like Pt distributions. Such morphology is consistent with limited Pt wetting and heterogeneous nucleation on carbon-based surfaces, where Pt nucleation is strongly influenced by defects, functional groups, and local surface chemistry.[28–30] The lack of a continuous Pt layer at this nominal thickness is therefore expected. The sputtered films (Fig. 2B) contained slightly larger and more densely distributed Pt islands than the evaporated films (Fig. 2A), which may reflect differences in the kinetic energy and flux of the incoming Pt species. However, the mentioned differences between these two samples are minor.

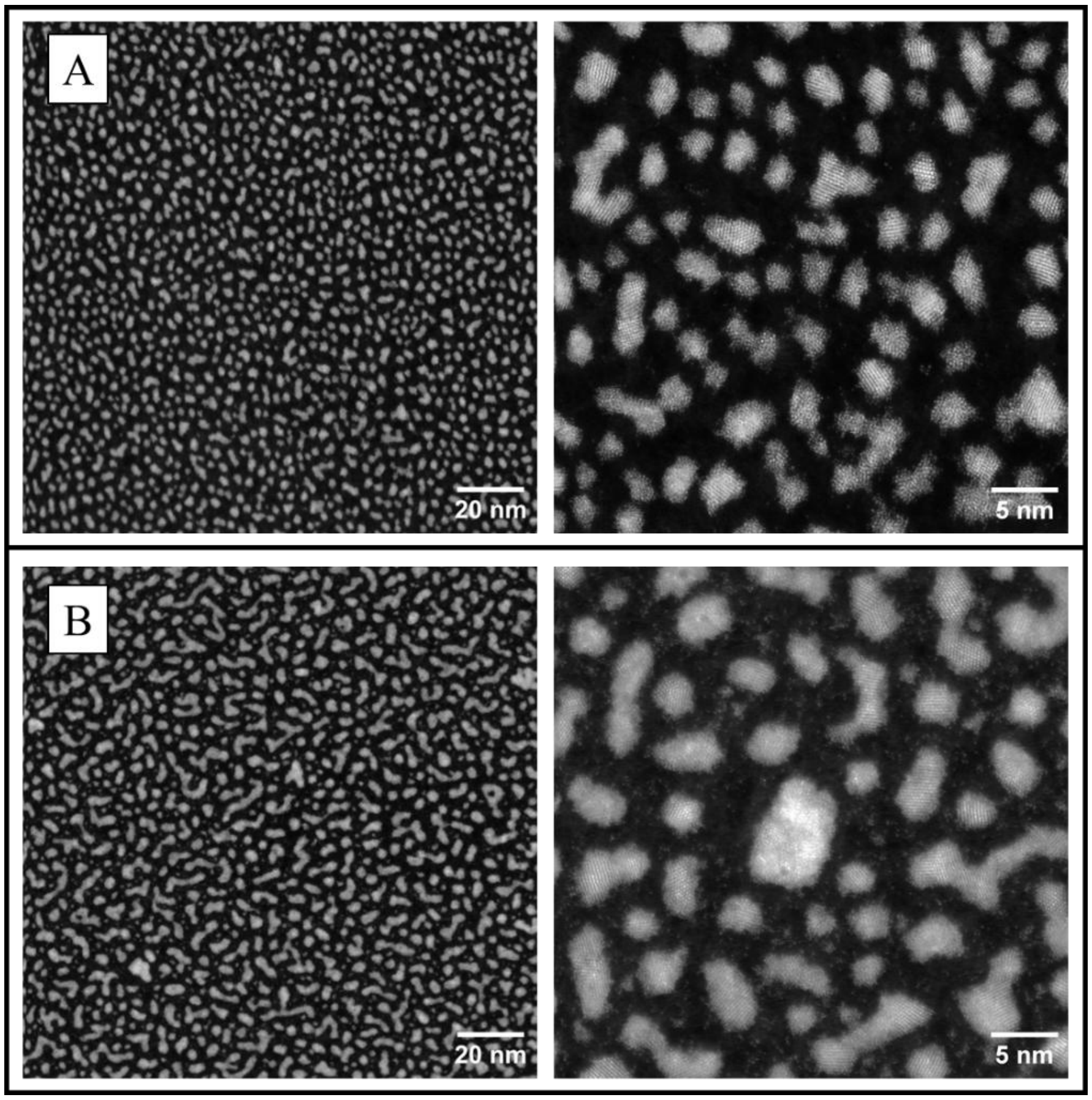

Figure 2 - HAADF STEM images of platinum films with a nominal thickness of 0.5 nm prepared by thermal evaporation (A) and magnetron sputtering (B)

After selenization, atomic resolution HAADF STEM images of rGO/$PtSe_2$ samples reveal an octahedral coordination of atoms, typical for 1T $PtSe_2$ [31] in both cases (Fig. 3). However, the island-like morphology of platinum is also reflected in grown $PtSe_2$ crystals. Samples produced from sputtered and evaporated platinum films both show a degree of inhomogeneity, though this time, the difference between the deposition techniques is more evident. While both layers exhibit clustered growth with overlapping parts appearing as brighter (high contrast) areas within the images, $PtSe_2$ grains prepared from sputtered platinum are connected, therefore covering a larger surface area. The in-plane lattice constant of the encapsulated $PtSe_2$ was determined from calibrated Fourier transforms (FTs) of HAADF STEM images to be 0.369 ± 0.002 nm, in good agreement with the literature values of 0.370–0.373 nm reported for monolayer $PtSe_2$. [6,32] An example of such an FT is shown in Figure 3D.

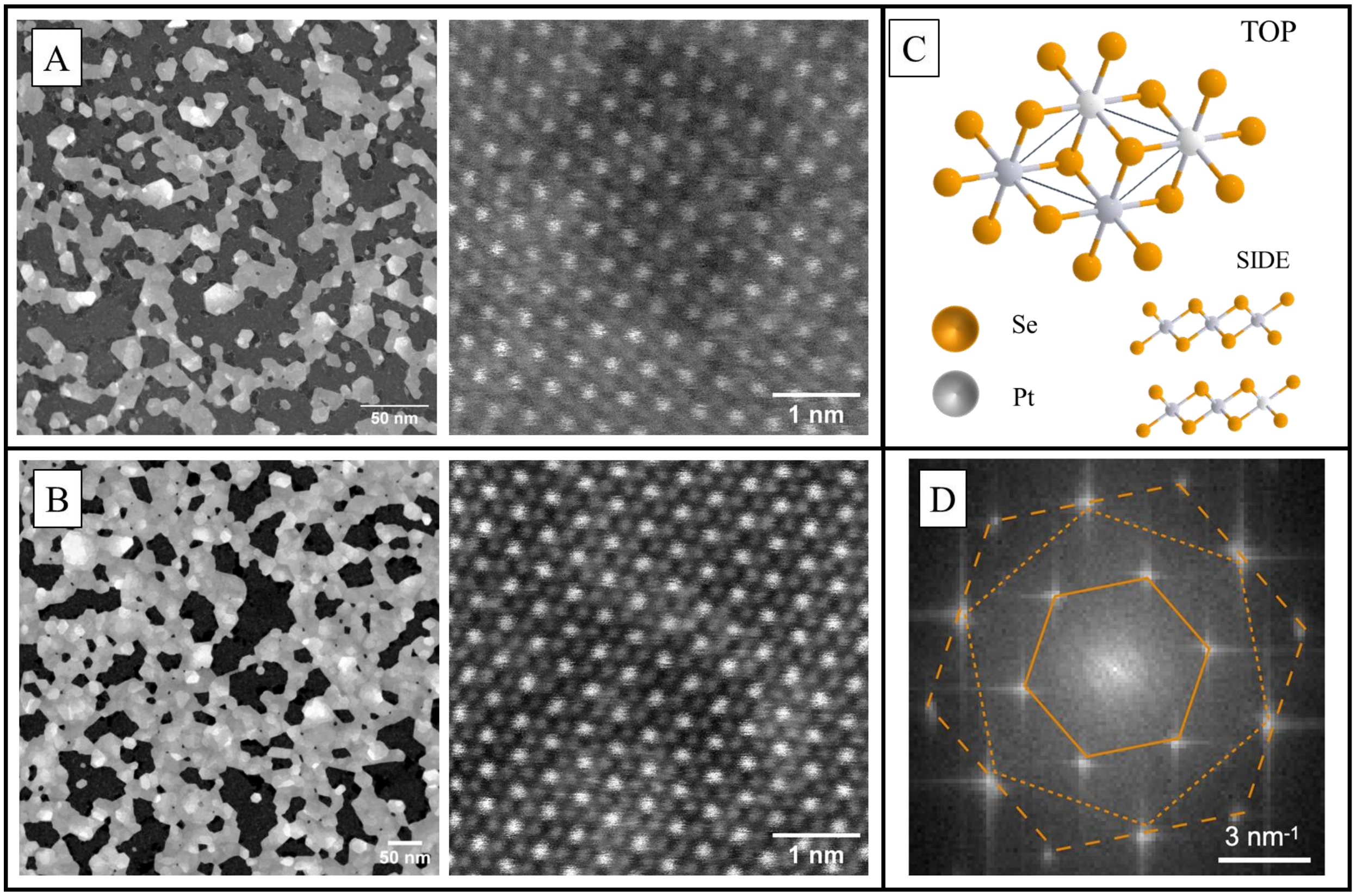

Figure 3 - HAADF STEM images of rGO/$PtSe_2$ heterostructures from platinum films with a nominal thickness of 0.5 nm prepared by thermal evaporation (A) and magnetron sputtering (B). Selenization conditions 550 °C/30 min. Structural model of 1T $PtSe_2$ (C). A calibrated Fourier transform showing the first three orders of $PtSe_2$ Bragg peaks (D).

Next, rGO/Pt/GO samples were examined after selenization. Figure 4 shows the fabrication of rGO/$PtSe_2$/rGO heterostructures in both cases of initial Pt layer used. In contrast to pristine graphene, GO contains oxygen functional groups, structural defects, flake boundaries, and interlayer nanochannels that can provide pathways for molecular transport.[33,34] During selenization, these features may allow selenium-containing species to access the underlying Pt layer through the GO overlayer. In addition to Se diffusion through the GO, selenium atoms can also diffuse through the gap between the edge of a GO flake and the metallic layer. It is estimated that oxygen containing functional groups of GO introduce an approximate addition of ~0.4 nm to the carbon backbone, which we estimate to be the gap width between the GO flakes and the underlying metal.[35] When the graphene oxide is reduced (due to the reducing atmosphere of Se vapor and high temperature), the gap eventually closes, and the graphene layer remaining on the top of $PtSe_2$ becomes impermeable, acting as a protective barrier against environmental degradation. Moreover, the rGO layer on top of the platinum film creates a confined space and promotes the lateral growth of $PtSe_2$. As can be seen from Figure 4, with graphene serving as an encapsulant, the gaps previously observed between clusters of the grown material are minimized and a continuous thin-film coverage over the given areas is observed. Similar results were observed for samples produced from both platinum deposition techniques. This confirms that the top graphene layer is a key component, facilitating the growth of continuously connected $PtSe_2$ layers from extremely thin platinum films. Note that the atomic structure of rGO layers is not directly visible in these images due to its low Z-contrast as compared to the $PtSe_2$ layers, but its presence can nevertheless be confirmed for instance in Fourier transforms (FT) of HAADF STEM images recorded from areas covered only by rGO (see Fig. S3 in Supplementary material). The non-encapsulated reference sample (Fig. S3A) contains two rGO monolayers, whereas the encapsulated sample (Fig. S3B) contains four rGO monolayers as a result of the additional GO coating prior to selenization. The corresponding graphitic scattering observed in both samples, therefore, supports the retention of the additional GO-derived carbon layer in the encapsulated heterostructure.

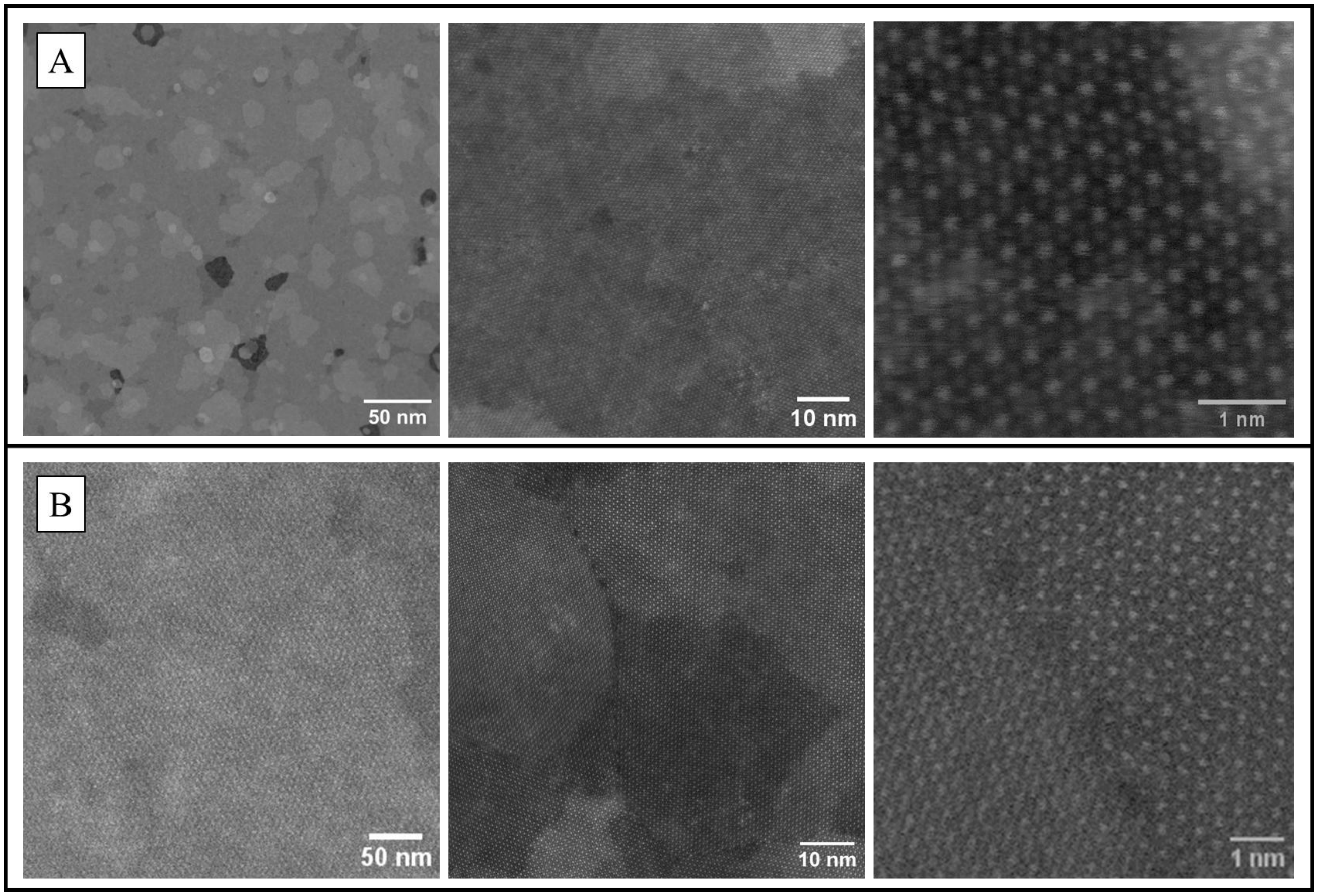

Figure 4 - HAADF STEM images of rGO/$PtSe_2$/rGO heterostructures from platinum films with a nominal thickness of 0.5 nm prepared by thermal evaporation (A) and magnetron sputtering (B). Selenization conditions 550 °C/30 min.

As noted previously, the final thickness of $PtSe_2$ grown from 0.5 nm thick Pt is up to 3 nm. [11] These unconfined reference films, however, exhibited more pronounced three-dimensional growth than the graphene-confined samples presented here. Because confinement promotes lateral spreading of the material, the $PtSe_2$ layers in the present heterostructures are expected to be thinner, with an estimated thickness of up to approximately six layers.

To verify the chosen selenization temperature of 550 °C discussed in the first paragraph of the results section, additional graphene-$PtSe_2$ sandwiches were prepared and selenized at 400 °C (Fig. 5). Herein, STEM imaging shows a clear misalignment of the material within the a-b axis, with multiple vertically oriented grains appearing as high-contrast lines by approximately 5 Å separation in the images. These observations indicate that while the rGO overlayer can influence the growth process, its presence alone is insufficient to achieve well-aligned $PtSe_2$. Selenization at elevated temperatures is additionally required to promote lateral crystal growth and suppress the formation of vertically oriented grains.

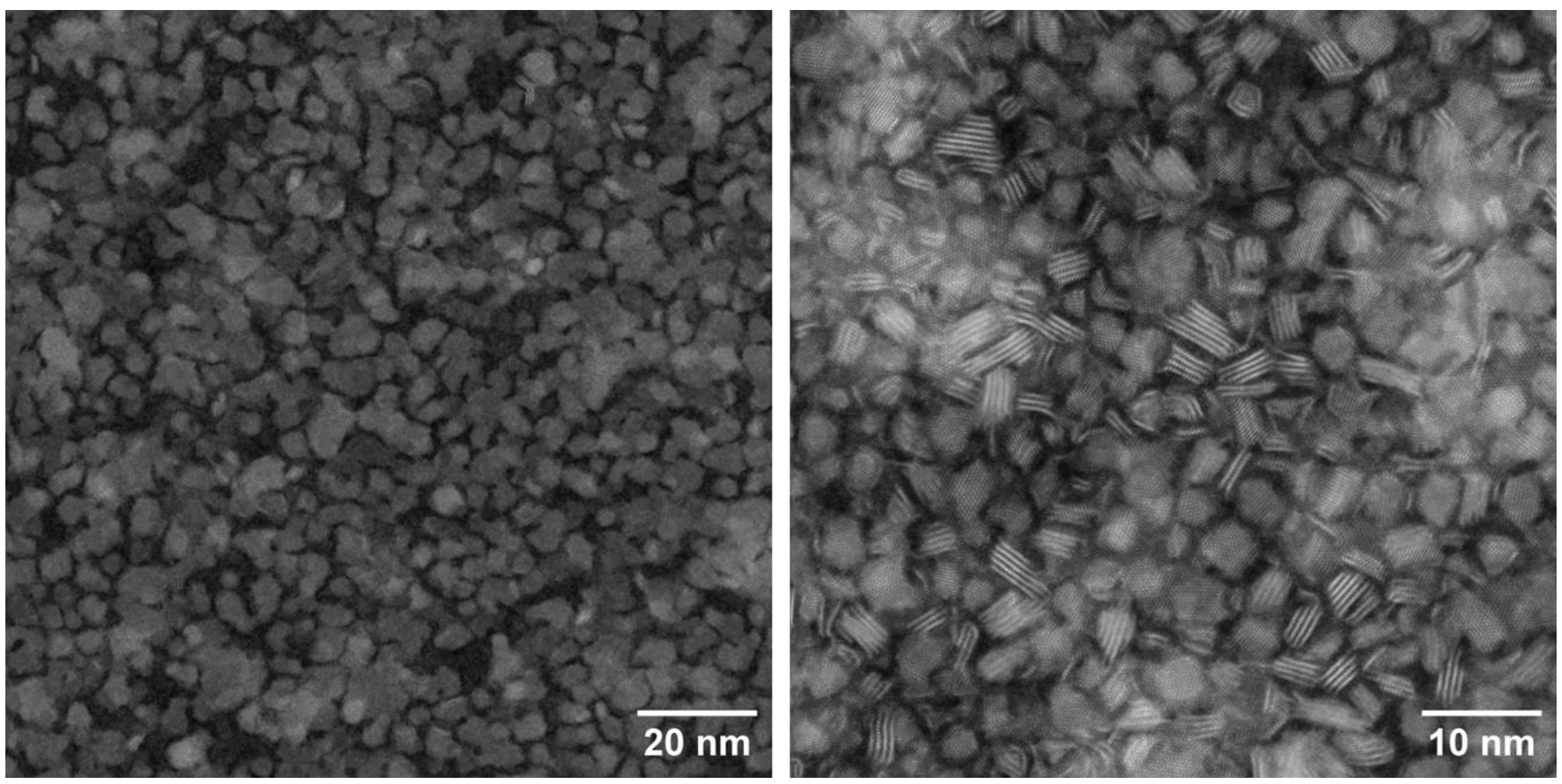

Figure 5 - HAADF STEM images of rGO/$PtSe_2$/rGO heterostructures from platinum films with a nominal thickness of 0.5 nm prepared by magnetron sputtering. Selenization conditions 400 °C/30 min.

To complement the local structural information obtained by STEM on SiN TEM grids, equivalent heterostructures were prepared on 1 × 1 $cm^2$ sapphire substrates using the same fabrication procedure. The larger substrate area enabled complementary characterization by X-ray diffraction, Raman spectroscopy, and X-ray photoelectron spectroscopy. These measurements were used to examine the structural and chemical characteristics of the $PtSe_2$/rGO and rGO/$PtSe_2$/rGO heterostructures and to assess their spatial homogeneity over length scales ranging from micrometers to centimeters.

Figure 6 shows the XRD patterns of the $PtSe_2$/rGO heterostructures grown on c-plane sapphire substrates and selenized at 400 and 550 °C. Both samples exhibit the $PtSe_2$ (001) reflection at approximately 2θ ~ 16.5°, confirming the formation of preferentially oriented layered $PtSe_2$ with basal planes parallel to the substrate surface. The higher intensity of the (001) reflection in the sample selenized at 550 °C indicates improved crystalline ordering compared with that of the heterostructure prepared at 400 °C. The weak features adjacent to the main $PtSe_2$ (001) peak arise from Laue oscillations, demonstrating coherent thickness and smooth interfaces of the layered $PtSe_2$ film. A broad, low-intensity feature centered near 2θ ~ 25° can be assigned to the graphitic (002) reflection of rGO.

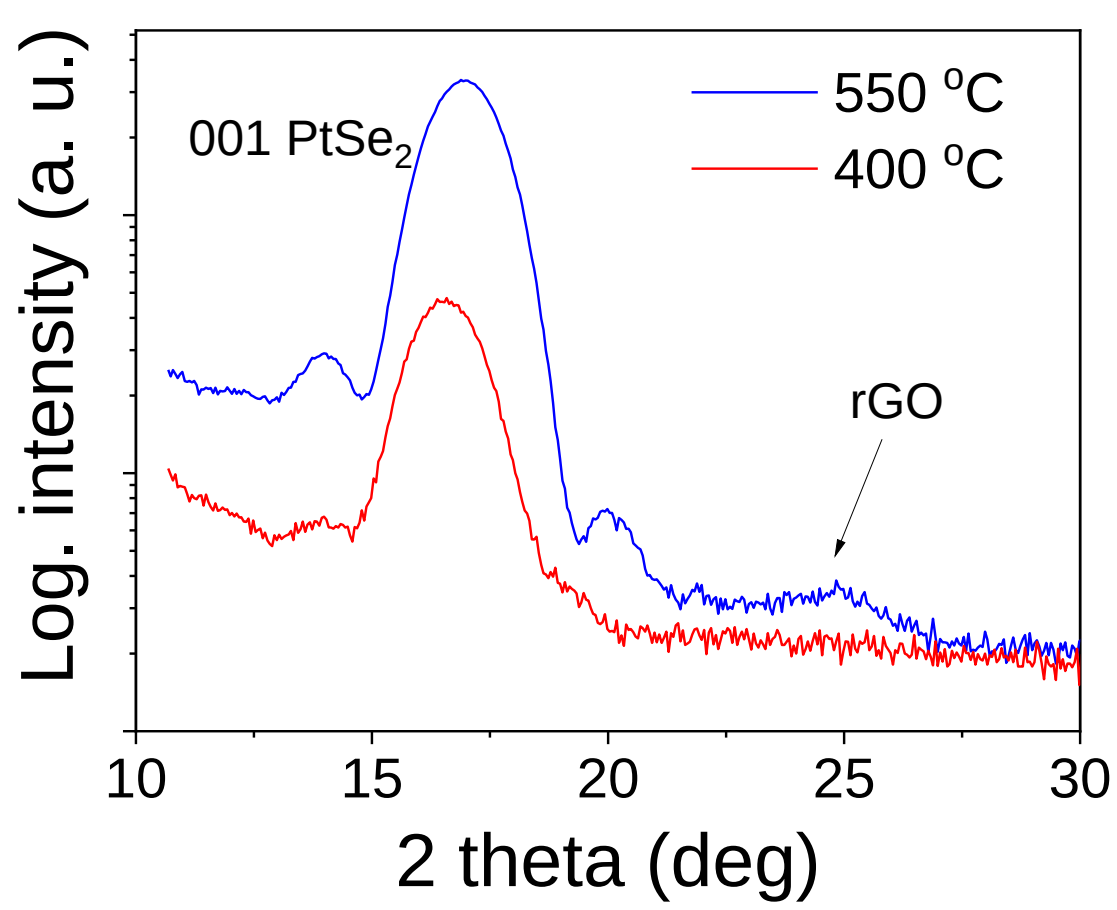

Figure 6 - XRD patterns of $PtSe_2$/rGO heterostructures grown on c-plane sapphire substrates and selenized at 400 and 550 °C. The $PtSe_2$ (001) reflection and adjacent Laue oscillations are indicated, together with the broad rGO-related (002) feature.

Raman spectroscopy was used to further characterize the $PtSe_2$/rGO heterostructures grown on c-plane sapphire substrates. The spectrum in Figure 7a shows the characteristic $PtSe_2$ Raman modes, including the dominant $E_g$ mode at approximately 176 $cm^{-1}$, together with the weaker $A_{1g}$ and longitudinal-optical (LO) modes at higher Raman shifts. The spectrum also displays the characteristic D and G bands of rGO at approximately 1350 and 1590 $cm^{-1}$, respectively, confirming the presence of a defective graphitic carbon component in the heterostructure.

Raman maps were acquired from the same region using the $PtSe_2$ $E_g$ mode and the rGO G band, as shown in Figure 7b, c. The $PtSe_2$-related signal was detected across the mapped area, while regions with reduced $PtSe_2$ intensity generally corresponded to areas of increased rGO-related intensity. The Raman spectra collected from these regions exhibit vibrational features attributable to both rGO and $PtSe_2$. However, the rGO signal may include contributions from the rGO layers located below and above the $PtSe_2$, and Raman spectroscopy alone cannot resolve these individual contributions or determine the vertical arrangement of the layers.

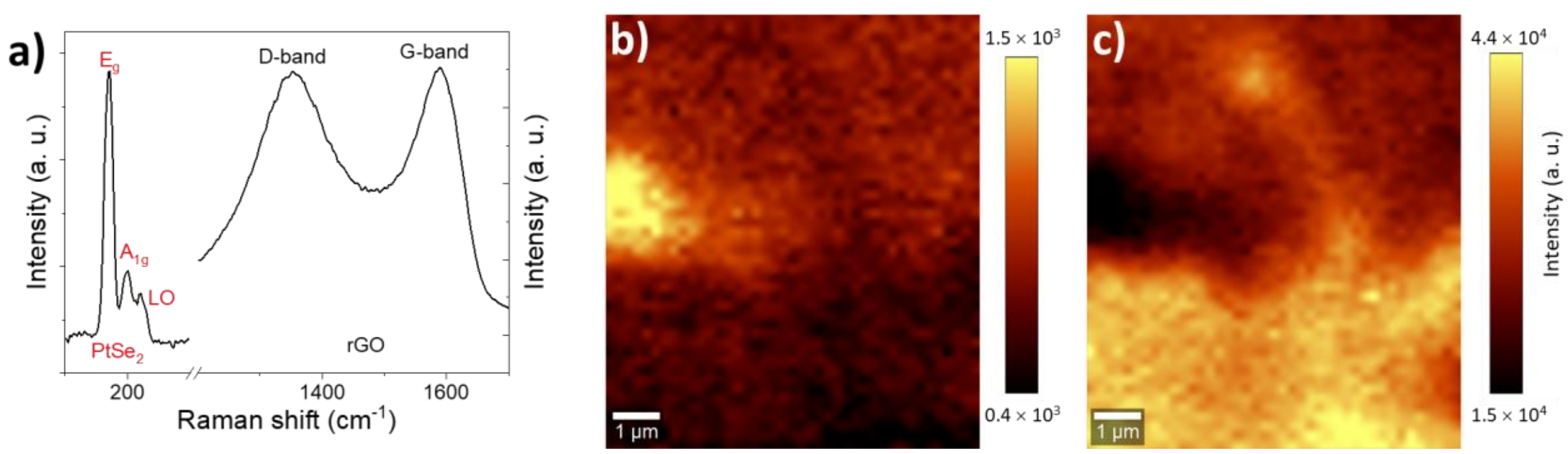

Figure 7 - Raman characterization of rGO/$PtSe_2$/rGO heterostructures grown on c-plane sapphire substrates. (a) Representative Raman spectrum showing the $PtSe_2$ $E_g$, $A_{1g}$, and LO modes, together with the rGO D and G bands. Raman maps (total area under a peak) acquired from the same area using the (b) $PtSe_2$ $E_g$ mode and (c) rGO G band.

The chemical state and composition were analyzed by XPS. Figure 8 shows Pt 4f, Se 3d, and C 1s spectra obtained from the $PtSe_2$/rGO heterostructure prepared by selenization of a pre-deposited Pt layer with a nominal thickness of 0.5 nm, coated with a thin GO layer. The Pt $4f_{7/2}$ and Se $3d_{5/2}$ peak maxima, measured at 73.0 ± 0.1 and 54.3 ± 0.1 eV, respectively, together with the Se stoichiometry of 1.9 ± 0.2, determined from the Se 3d and Pt 4f XPS intensities, and the weak signals from off-stoichiometric $Pt_xSe_y$, confirm the formation of near-stoichiometric $PtSe_2$. [11] The spectrum in Fig. 8c displays the C 1s spectrum of the heterostructure. The dominant C 1s component, measured at 284.5 ± 0.1 eV, corresponds to carbon atoms in $sp^2$ hybridization [36–38] and confirms the reduction of the GO coating to a graphene-based carbon overlayer after the selenization step. It is worth noting that under the selenization conditions employed in this study, complete selenization of metallic platinum may not occur with a thicker GO coating (see Fig. S4 in the Supplementary Material for further details). The trade-off between the selenium transport across the GO layer and the rate of chemical reduction of oxide groups in GO at the elevated temperatures would set a limit to selenization process, because a resulting rGO membrane acts as a tight molecular barrier, which would prevent selenium to cross through. The determination of optimal conditions for the complete selenization of platinum coated with GO of varying thicknesses falls outside the scope of this study and will be explored in future research.

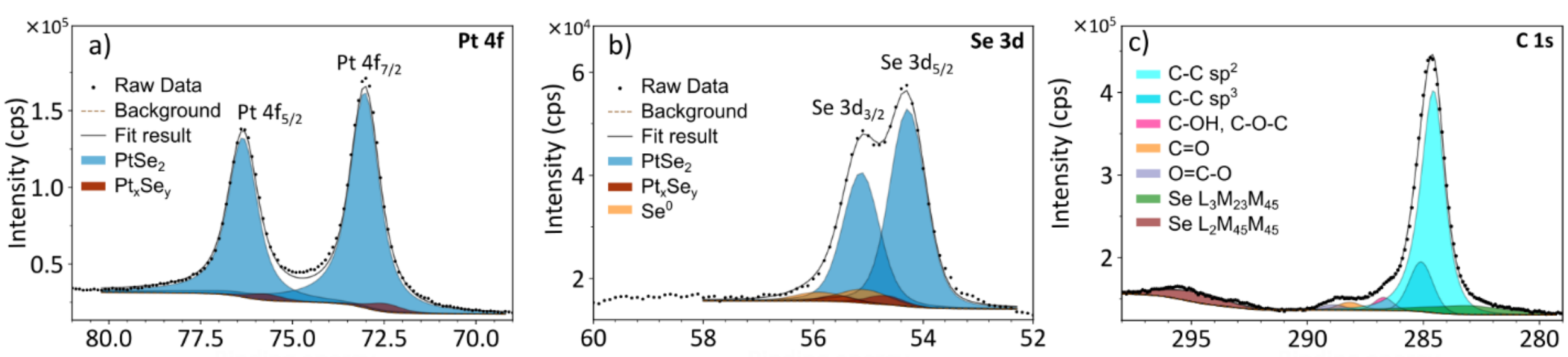

Figure 8 – XPS spectra of $PtSe_2$/rGO heterostructure prepared on *c*-plane sapphire substrate. (a) Pt 4f, (b) Se 3d, and (c) C 1s core level spectra, and their decomposition into components.

Lastly, in situ heating experiments were performed to evaluate the thermal stability of the rGO/$PtSe_2$/rGO heterostructures (Fig. 9) under UHV conditions in the STEM. Plan-view HAADF-STEM images revealed no apparent changes in grain size or spatial distribution during annealing from below the growth temperature (400 °C) up to 1200 °C, the highest temperature accessible with the Protochips Fusion holders. These observations demonstrate morphological persistence of the encapsulated heterostructures under the specific in situ UHV conditions. However, plan-view STEM imaging alone cannot verify the retention of $PtSe_2$ stoichiometry, selenium content, phase composition, or crystalline quality. Therefore, no conclusions can be drawn regarding the chemical stability of $PtSe_2$ or the effectiveness of the rGO overlayer in suppressing selenium desorption or preventing thermal degradation. Nevertheless, the observed morphological persistence indicates that the rGO/$PtSe_2$/rGO architecture remains structurally intact across this broad temperature range under the conditions investigated.

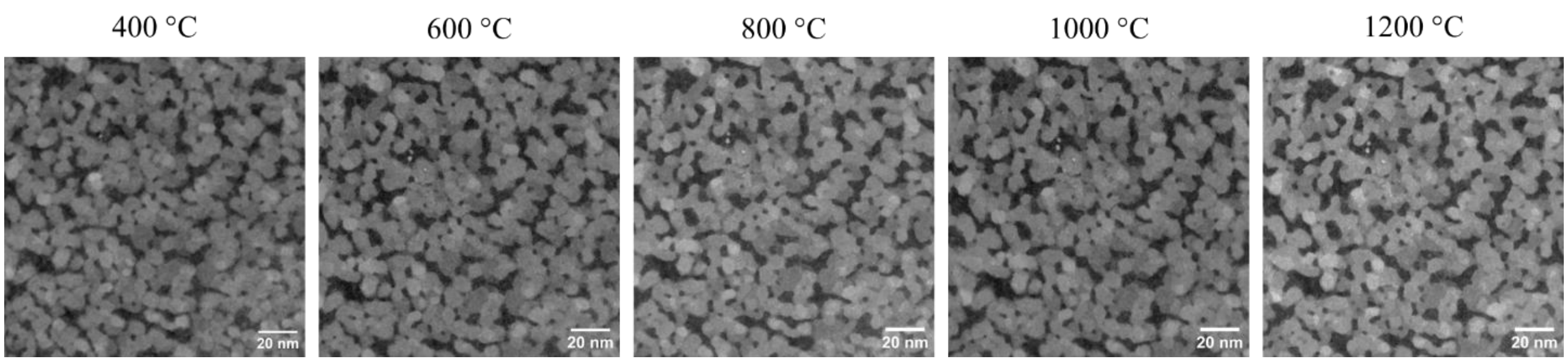

Figure 9 - Thermal stability experiment – HAADF STEM images of in situ continuous annealing of rGO/$PtSe_2$/rGO heterostructures from platinum films with a nominal thickness of 0.5 nm prepared by magnetron sputtering. Selenization conditions 400 °C/30 min.

## CONCLUSION

While platinum and its dichalcogenides are relatively stable under ambient conditions, many other transition metal dichalcogenides are susceptible to oxidation, which can rapidly alter their electronic properties. We propose that our method of graphene sandwiching will not only facilitate the formation of ultrathin TMD layers but also be a valuable method for the fabrication of compounds unstable in the environment, allowing a more thorough investigation of their properties and possible integration into electronic components.

In this work, we investigated a graphene-mediated confinement strategy for the synthesis of few-layer $PtSe_2$ films through the selenization of Pt layers coated with graphene oxide. The resulting $PtSe_2$ morphology was governed by the combined effects of the graphene-based overlayer and the selenization temperature. Under the selected selenization conditions, graphene-mediated confinement limited the formation of large three-dimensional $PtSe_2$ features and favored lateral growth of $PtSe_2$ within the rGO-based architecture. We showed that elevated selenization temperatures are essential for promoting lateral crystal growth and improving structural alignment in the rGO/$PtSe_2$ heterostructure.

The presented approach addresses one of the key challenges in $PtSe_2$ synthesis, namely the controlled fabrication of ultrathin films with high lateral continuity. By combining confinement-assisted growth with direct structural observation at the atomic scale, this study provides new insight into the mechanisms governing TMD formation under constrained growth conditions. Temperature-resolved in-situ STEM investigations showed no obvious morphological changes in the encapsulated structures during heating under the investigated UHV conditions.

The demonstrated confinement-assisted growth mechanism may be applicable beyond the Pt–Se system and provide a general route for the fabrication of ultrathin layered materials under vertically constrained growth conditions. By simultaneously controlling film morphology and protecting the growing material from the environment, graphene sandwiching offers a versatile platform for the synthesis and investigation of low-dimensional compounds that are difficult to realize using conventional approaches.

## Acknowledgements

This work was supported by the Slovak Research and Development Agency, APVV SK-AT-23-0021, APVV-23-0564 and Slovak Grant Agency for Science, VEGA 2/0075/25. K.M. acknowledges funding from OeAD and FWF through grants SK09/2024 and P35912. A. L. acknowledges funding from project "Ultra-thin (few-layer) TMD films for electronic applications" under the SAS Return Project Scheme.

## References

[1] F. Zheng, W. Meng, L.-J. Li, Continue the Scaling of Electronic Devices with Transition Metal Dichalcogenide Semiconductors, Nano Lett. 25 (2025) 3683–3691. https://doi.org/10.1021/acs.nanolett.4c06007.

[2] T. Lu, Y. Wang, G. Cai, H. Jia, X. Liu, C. Zhang, S. Meng, M. Liu, Synthesizability of transition-metal dichalcogenides: a systematic first-principles evaluation, Mater. Futur. 2 (2023) 015001. https://doi.org/10.1088/2752-5724/acbe10.

[3] V. Sorkin, H. Pan, H. Shi, S.Y. Quek, Y.W. Zhang, Nanoscale Transition Metal Dichalcogenides: Structures, Properties, and Applications, Crit. Rev. Solid State Mater. Sci. 39 (2014) 319–367. https://doi.org/10.1080/10408436.2013.863176.

[4] A. Kuc, T. Heine, The electronic structure calculations of two-dimensional transition-metal dichalcogenides in the presence of external electric and magnetic fields, Chem. Soc. Rev. 44 (2015) 2603–2614. https://doi.org/10.1039/C4CS00276H.

[5] Y. Gong, Z. Lin, Y.-X. Chen, Q. Khan, C. Wang, B. Zhang, G. Nie, N. Xie, D. Li, Two-Dimensional Platinum Diselenide: Synthesis, Emerging Applications, and Future Challenges, Nano-Micro Lett. 12 (2020) 174. https://doi.org/10.1007/s40820-020-00515-0.

[6] A. Kandemir, B. Akbali, Z. Kahraman, S.V. Badalov, M. Ozcan, F. Iyikanat, H. Sahin, Structural, electronic and phononic properties of $PtSe_2$ : from monolayer to bulk, Semicond. Sci. Technol. 33 (2018) 085002. https://doi.org/10.1088/1361-6641/aacba2.

[7] S. Daws, P. Kotak, C.-N. Kuo, C. Shan Lue, A. Politano, C. Lamuta, Platinum diselenide PtSe2: An ambient-stable material for flexible electronics, Mater. Sci. Eng. B 283 (2022) 115824. https://doi.org/10.1016/j.mseb.2022.115824.

[8] Z. Huang, W. Zhang, W. Zhang, Computational Search for Two-Dimensional MX2 Semiconductors with Possible High Electron Mobility at Room Temperature, Materials 9 (2016) 716. https://doi.org/10.3390/ma9090716.

[9] W. Zhang, Z. Huang, W. Zhang, Y. Li, Two-dimensional semiconductors with possible high room temperature mobility, Nano Res. 7 (2014) 1731–1737. https://doi.org/10.1007/s12274-014-0532-x.

[10] L. Ansari, S. Monaghan, N. McEvoy, C.Ó. Coileáin, C.P. Cullen, J. Lin, R. Siris, T. Stimpel-Lindner, K.F. Burke, G. Mirabelli, R. Duffy, E. Caruso, R.E. Nagle, G.S. Duesberg, P.K. Hurley, F. Gity, Quantum confinement-induced semimetal-to-semiconductor evolution in

large-area ultra-thin PtSe2 films grown at 400 °C, Npj 2D Mater. Appl. 3 (2019) 33. https://doi.org/10.1038/s41699-019-0116-4.
[11] M. Sojková, E. Dobročka, P. Hutár, V. Tašková, L. Pribusová Slušná, R. Stoklas, I. Píš, F. Bondino, F. Munnik, M. Hulman, High carrier mobility epitaxially aligned PtSe2 films grown by one-zone selenization, Appl. Surf. Sci. 538 (2021) 147936. https://doi.org/10.1016/j.apsusc.2020.147936.
[12] Y. Zhao, J. Qiao, Z. Yu, P. Yu, K. Xu, S.P. Lau, W. Zhou, Z. Liu, X. Wang, W. Ji, Y. Chai, High-Electron-Mobility and Air-Stable 2D Layered $PtSe_2$ FETs, Adv. Mater. 29 (2017) 1604230. https://doi.org/10.1002/adma.201604230.
[13] H. Ma, Q. Qian, B. Qin, Z. Wan, R. Wu, B. Zhao, H. Zhang, Z. Zhang, J. Li, Z. Zhang, B. Li, L. Wang, X. Duan, Controlled Synthesis of Ultrathin $PtSe_2$ Nanosheets with Thickness-Tunable Electrical and Magnetoelectrical Properties, Adv. Sci. 9 (2022) 2103507. https://doi.org/10.1002/advs.202103507.
[14] M. Tharrault, E. Desgué, D. Carisetti, B. Plaçais, C. Voisin, P. Legagneux, E. Baudin, Raman spectroscopy of monolayer to bulk $PtSe_2$ exfoliated crystals, 2D Mater. 11 (2024) 025011. https://doi.org/10.1088/2053-1583/ad1e79.
[15] A.V. Kolobov, J. Tominaga, Two-Dimensional Transition-Metal Dichalcogenides, Springer International Publishing, Cham, 2016. https://doi.org/10.1007/978-3-319-31450-1.
[16] M. Sojková, J. Hrdá, S. Volkov, K. Vegso, A. Shaji, T. Vojteková, L.P. Slušná, N. Gál, E. Dobročka, P. Siffalovic, T. Roch, M. Gregor, M. Hulman, Growth of PtSe2 few-layer films on NbN superconducting substrate, Appl. Phys. Lett. 119 (2021) 013101. https://doi.org/10.1063/5.0053309.
[17] L. Chen, Z. Cheng, S. He, X. Zhang, K. Deng, D. Zong, Z. Wu, M. Xia, Large-area single-crystal TMD growth modulated by sapphire substrates, Nanoscale 16 (2024) 978–1004. https://doi.org/10.1039/D3NR05400D.
[18] Z. Li, S. Wu, W. Lv, J. Shao, F. Kang, Q. Yang, Graphene Emerges as a Versatile Template for Materials Preparation, Small 12 (2016) 2674–2688. https://doi.org/10.1002/smll.201503722.
[19] D. Moon, W. Lee, C. Lim, J. Kim, J. Kim, Y. Jung, H.-Y. Choi, W.S. Choi, H. Kim, J.-H. Baek, C. Kim, J. Joo, H.-G. Oh, H. Jang, K. Watanabe, T. Taniguchi, S. Bae, J. Son, H. Ryu, J. Kwon, H. Cheong, J.W. Han, H. Jang, G.-H. Lee, Hypotaxy of wafer-scale single-crystal transition metal dichalcogenides, Nature 638 (2025) 957–964. https://doi.org/10.1038/s41586-024-08492-9.
[20] K. Mustonen, C. Hofer, P. Kotrusz, A. Markevich, M. Hulman, C. Mangler, T. Susi, T.J. Pennycook, K. Hricovini, C. Richter, J.C. Meyer, J. Kotakoski, V. Skákalová, Toward Exotic Layered Materials: 2D Cuprous Iodide, Adv. Mater. 34 (2022) 2106922. https://doi.org/10.1002/adma.202106922.
[21] A.K. Geim, Graphene: Status and Prospects, Science 324 (2009) 1530–1534. https://doi.org/10.1126/science.1158877.
[22] J. Azadmanjiri, V.K. Srivastava, P. Kumar, Z. Sofer, J. Min, J. Gong, Graphene-Supported 2D transition metal dichalcogenide van der waals heterostructures, Appl. Mater. Today 19 (2020) 100600. https://doi.org/10.1016/j.apmt.2020.100600.
[23] T. Han, H. Liu, S. Wang, S. Chen, K. Yang, Research on the Preparation and Spectral Characteristics of Graphene/TMDs Hetero-structures, Nanoscale Res. Lett. 15 (2020) 219. https://doi.org/10.1186/s11671-020-03439-1.
[24] G. Kerherve, W.S.J. Skinner, J.A. Hochhaus, A. Graf, D.J. Morgan, M.A. Isaacs, B.P. Reed, H. Nakajima, D.J. Payne, KherveFitting: An Open Source Software for Fitting X-Ray Photoelectron Spectroscopy Data, Surf. Interface Anal. 58 (2026) 54–65. https://doi.org/10.1002/sia.70032.
[25] M. Sojková, O. Pohorelec, J. Hrdá, T.E. Krajčovičová, A. Kozak, J. Dérer, M. Hulman, M. Ťapajna, S. Egorov, P. Kotrusz, V. Skákalová, K. Mustonen, F. Bondino, I. Píš, D.

Gregušová, Scalable transfer-free fabrication of PtSe2 FETs, FlatChem 59 (2026) 101091. https://doi.org/10.1016/j.flatc.2026.101091.
[26] M.S. Shawkat, J. Gil, S.S. Han, T.-J. Ko, M. Wang, D. Dev, J. Kwon, G.-H. Lee, K.H. Oh, H.-S. Chung, T. Roy, Y. Jung, Y. Jung, Thickness-Independent Semiconducting-to-Metallic Conversion in Wafer-Scale Two-Dimensional $PtSe_2$ Layers by Plasma-Driven Chalcogen Defect Engineering, ACS Appl. Mater. Interfaces 12 (2020) 14341–14351. https://doi.org/10.1021/acsami.0c00116.
[27] J. Ahn, I. Hong, G. Lee, H. Shin, A. Benali, Y. Kwon, J.T. Krogel, Stacking polymorphism of PtSe2: its implication to layer-dependent metal-insulator transitions, Npj 2D Mater. Appl. 9 (2025) 34. https://doi.org/10.1038/s41699-025-00544-6.
[28] B. Karasulu, R.H.J. Vervuurt, W.M.M. Kessels, A.A. Bol, Continuous and ultrathin platinum films on graphene using atomic layer deposition: a combined computational and experimental study, Nanoscale 8 (2016) 19829–19845. https://doi.org/10.1039/C6NR07483A.
[29] M.W. Iqbal, M.Z. Iqbal, M.F. Khan, X. Jin, C. Hwang, J. Eom, Modification of the structural and electrical properties of graphene layers by Pt adsorbates, Sci. Technol. Adv. Mater. 15 (2014) 055002. https://doi.org/10.1088/1468-6996/15/5/055002.
[30] K. Kim, H.-B.-R. Lee, R.W. Johnson, J.T. Tanskanen, N. Liu, M.-G. Kim, C. Pang, C. Ahn, S.F. Bent, Z. Bao, Selective metal deposition at graphene line defects by atomic layer deposition, Nat. Commun. 5 (2014) 4781. https://doi.org/10.1038/ncomms5781.
[31] G.H. Ryu, J. Chen, Y. Wen, J.H. Warner, In-Situ Atomic-Scale Dynamics of Thermally Driven Phase Transition of 2D Few-Layered 1T $PtSe_2$ into Ultrathin 2D Nonlayered PtSe Crystals, Chem. Mater. 31 (2019) 9895–9903. https://doi.org/10.1021/acs.chemmater.9b04274.
[32] D.A. Chenet, B. Aslan, P.Y. Huang, C. Fan, A.M. Van Der Zande, T.F. Heinz, J.C. Hone, In-Plane Anisotropy in Mono- and Few-Layer $ReS_2$ Probed by Raman Spectroscopy and Scanning Transmission Electron Microscopy, Nano Lett. 15 (2015) 5667–5672. https://doi.org/10.1021/acs.nanolett.5b00910.
[33] T. Yang, H. Lin, K.P. Loh, B. Jia, Fundamental Transport Mechanisms and Advancements of Graphene Oxide Membranes for Molecular Separation, Chem. Mater. 31 (2019) 1829–1846. https://doi.org/10.1021/acs.chemmater.8b03820.
[34] H. Yang, J. Baek, H.G. Park, Architecture and mass transport properties of graphene-based membranes, JMST Adv. 2 (2020) 77–88. https://doi.org/10.1007/s42791-020-00032-6.
[35] H.C. Schniepp, J.-L. Li, M.J. McAllister, H. Sai, M. Herrera-Alonso, D.H. Adamson, R.K. Prud'homme, R. Car, D.A. Saville, I.A. Aksay, Functionalized Single Graphene Sheets Derived from Splitting Graphite Oxide, J. Phys. Chem. B 110 (2006) 8535–8539. https://doi.org/10.1021/jp060936f.
[36] A. Kovtun, D. Jones, S. Dell'Elce, E. Treossi, A. Liscio, V. Palermo, Accurate chemical analysis of oxygenated graphene-based materials using X-ray photoelectron spectroscopy, Carbon 143 (2019) 268–275. https://doi.org/10.1016/j.carbon.2018.11.012.
[37] V. Skákalová, P. Kotrusz, M. Jergel, T. Susi, A. Mittelberger, V. Vretenár, P. Šiffalovič, J. Kotakoski, J.C. Meyer, M. Hulman, Chemical Oxidation of Graphite: Evolution of the Structure and Properties, J. Phys. Chem. C 122 (2018) 929–935. https://doi.org/10.1021/acs.jpcc.7b10912.
[38] A. Carvalho, M.C.F. Costa, V.S. Marangoni, P.R. Ng, T.L.H. Nguyen, A.H. Castro Neto, The Degree of Oxidation of Graphene Oxide, Nanomaterials 11 (2021) 560. https://doi.org/10.3390/nano11030560.

# Supplementary material for

## Graphene-Confined Growth of Few-Layer $PtSe_2$ via One-Step Selenization

Timea Ema Krajčovičová,[1] Peter Kotrusz,[1,2] Viera Skákalová,[1,2] Agáta Laurenčíková,[1] Igor Píš,[1,3] Martin Hulman,[1,2] Kimmo Mustonen [3] and Michaela Sojková,[1*]

[1] *Institute of Electrical Engineering, Slovak Academy of Sciences, Dúbravská cesta 9, 84104 Bratislava, Slovakia*

[2] *Danubia NanoTech, s.r.o., Dúbravská cesta 9, 84104 Bratislava, Slovakia*

[3] *Polymer Institute, Slovak Academy of Sciences, Dúbravská cesta 9, 845 41 Bratislava, Slovakia*

[4] *University of Vienna, Faculty of Physics, Boltzmanngasse 5, 1090 Vienna, Austria*

**Corresponding author:**

email: michaela.sojkova@savba.sk

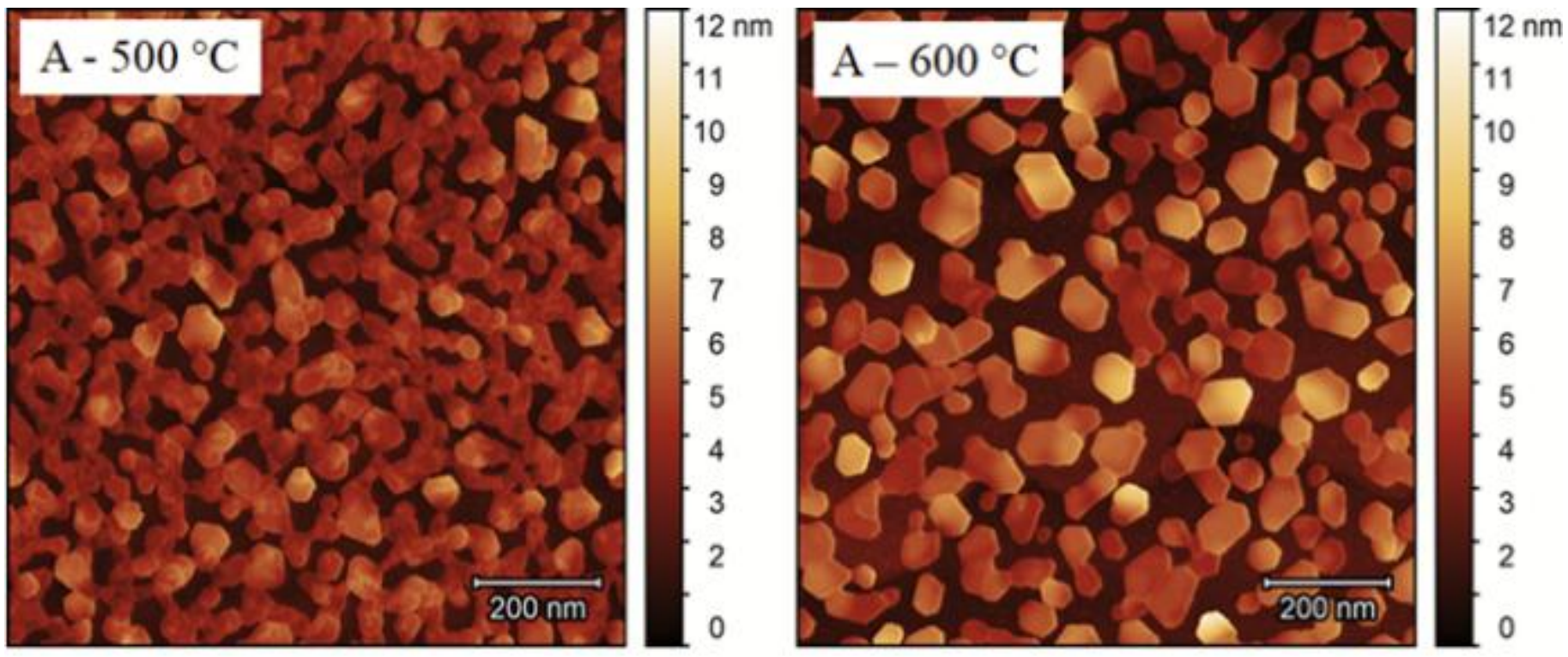

Figure S1. AFM topography images of $PtSe_2$ grown from 0.5 nm thick Pt layers on c-plane sapphire substrate after selenization at (a) 500 °C and (b) 600 °C. The maps show a discontinuous nanostructured surface with local topographic height variations. The color scale represents local AFM height, including isolated elevated features, and is not equivalent to the area-averaged effective film thickness determined by X-ray reflectivity.

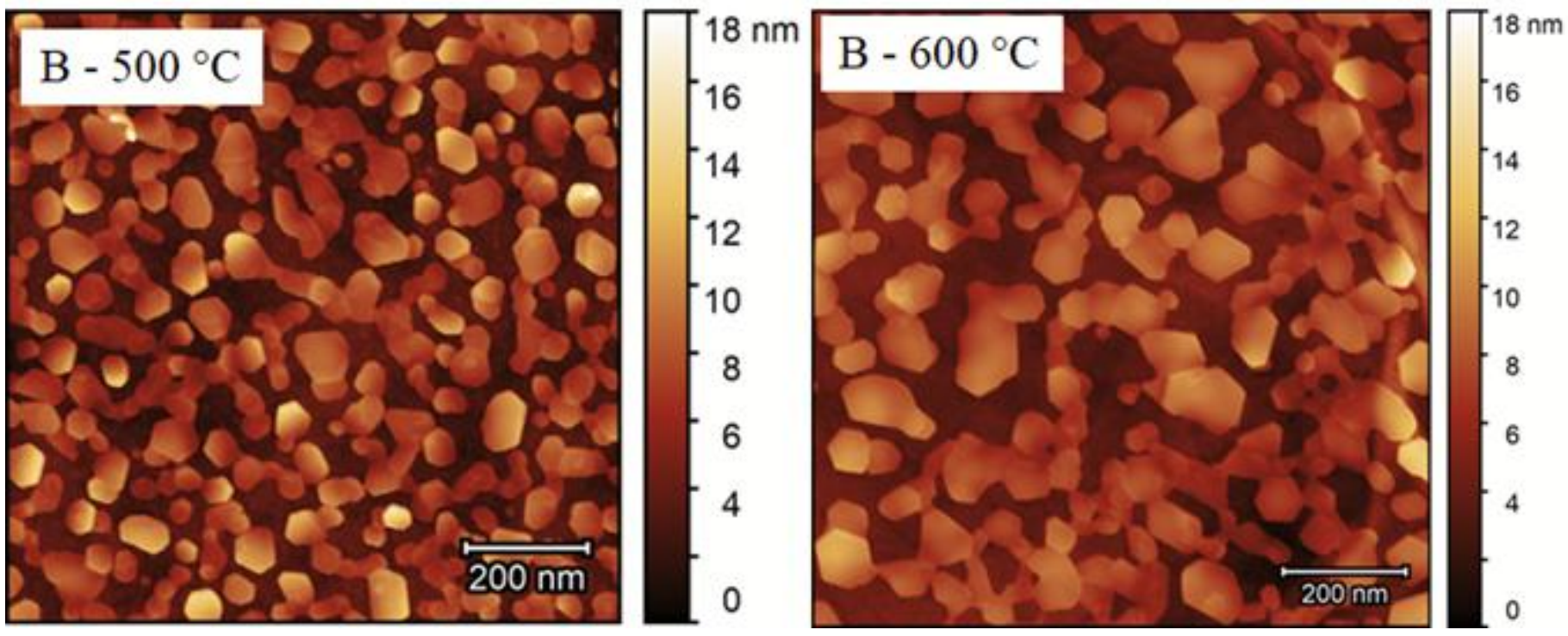

Figure S2. AFM topography images of $PtSe_2$ grown by selenization of nominally 0.5 nm thick Pt precursor layers on an rGO substrate at (a) 500 °C and (b) 600 °C. The images show local surface-height variations associated with the nanostructured $PtSe_2$/rGO morphology; the AFM height scale reflects local topographic features and does not represent the area-averaged effective film thickness.

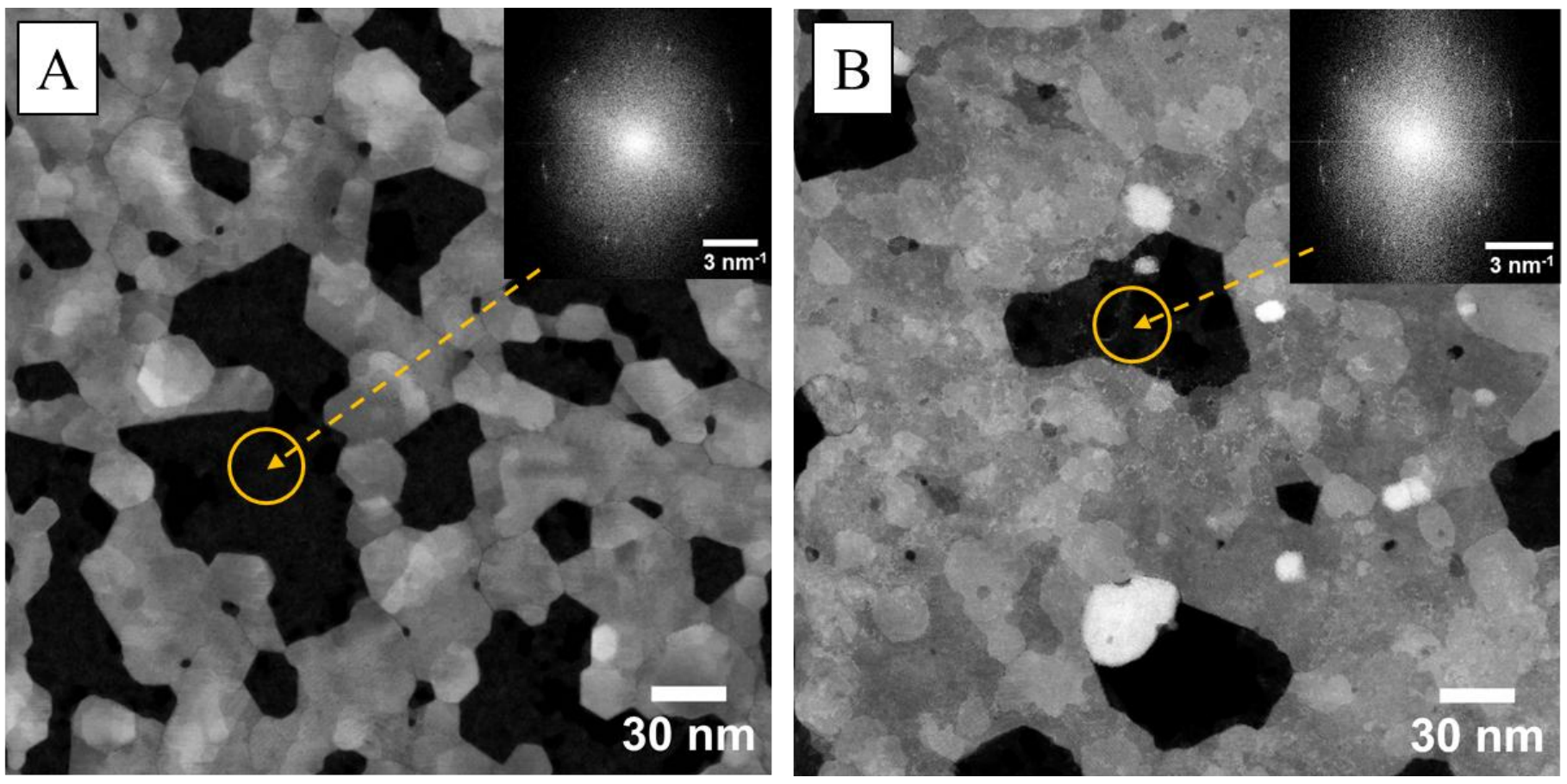

Figure S3 - Plan-view HAADF-STEM images of $PtSe_2$ on two rGO monolayers (A) and of the encapsulated heterostructure containing four rGO monolayers (B). Insets show Fourier transforms from $PtSe_2$-free regions marked by arrows, confirming the graphitic carbon layers.

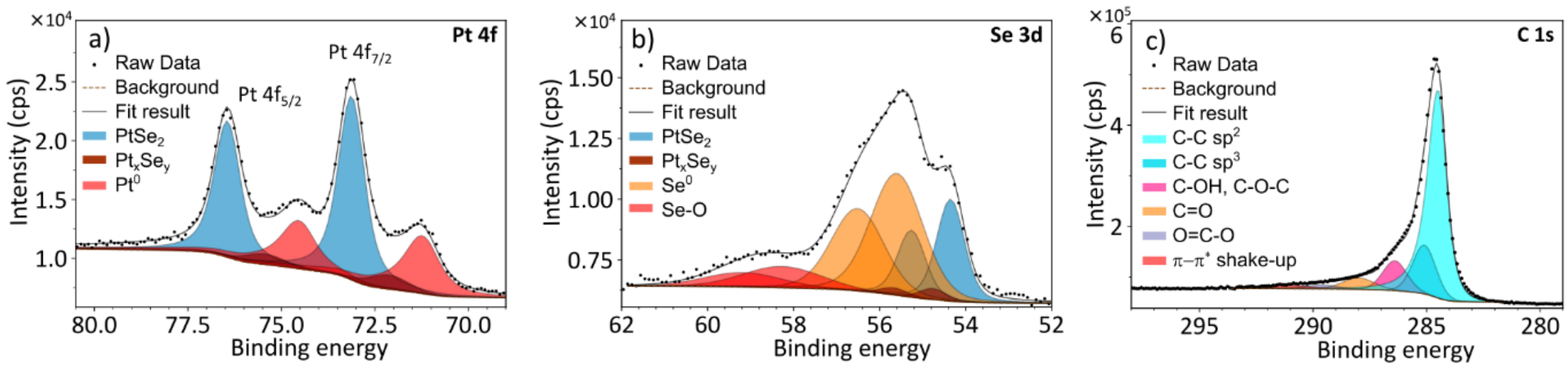

Figure S4 - XPS spectra of the $PtSe_2$/rGO heterostructure prepared on a c-plane sapphire substrate from a thicker GO coating. (a) Pt 4f, (b) Se 3d, and (c) C 1s core-level spectra, with decomposition into components.

Table S1. Composition of $PtSe_2$/rGO samples determined from XPS.

| Sample | Apparent surface chemical composition (at. %) | | | | | | | | | |
|---|---|---|---|---|---|---|---|---|---|---|
| | Pt $4f_{7/2}$ | | | Se $3d_{5/2}$ | | | | C 1s | O 1s | Al 2s |
| | **$PtSe_2$** (73.0 eV) | **$Pt_xSe_y$** (72.0 eV) | **$Pt^0$** (71.1 eV) | **$PtSe_2$** (54.3 eV) | **$Pt_xSe_y$,** (54.7 eV) | **$Se^0$** (55.4 eV) | **Se-O** (58.3 eV) | | | |
| **$PtSe_2$/rGO thinner** | 2.45 | 0.1 | 0 | 4.40 | 0.2 | 0.4 | 0 | 71.3 | 19.4 | 1.8 |
| **$PtSe_2$/rGO thicker** | 0.19 | 0.02 | 0.11 | 0.38 | 0.04 | 0.79 | 0.2 | 85.5 | 12.8 | 0 |

Figure S4 displays the Pt 4f, Se 3d, and C 1s spectra acquired from the selenized Pt layer with a nominal thickness of 0.5 nm, coated with a GO layer thicker than that described in the main text. The increased thickness is indicated by a lower apparent Pt concentration (Table S1), due to greater attenuation of signals from beneath the thicker rGO overlayer. In the Pt 4f and Se 3d spectra, in addition to $PtSe_2$-related components, signals from unreacted metallic platinum (Pt $4f_{7/2}$ peak at 71.1 ± 0.1 eV) and unreacted elemental Se (Se $3d_{5/2}$ peak at 55.4 ± 0.3 eV) were detected.